\documentclass[letterpaper,twocolumn,10pt]{article}
\usepackage{usenix-2020-09}

\usepackage{tikz,hyperref}
\usepackage{amsmath,comment,sansmath,flushend}
\usepackage[T1]{fontenc}
\usepackage{caption, subcaption, tikz, yfonts, booktabs,multirow,xspace,enumitem}
\usepackage[multiple]{footmisc}
\usepackage{filecontents}

\newcommand{\DeLSM}{DeLSM\xspace}

\newcommand*\circled[1]{\tikz[baseline=(char.base)]{
		\node[shape=circle,draw,inner sep=1pt] (char) {#1};}}	
\begin{document}

\date{}

 
\title{\Large \bf Accelerating LSM-Tree with the {\tt Dentry} Management of File System}


\author{
{\rm Yanpeng Hu and Chundong Wang\thanks{Corresponding Author (cd\_wang@outlook.com)}}\\
School of Information Science and Technology\\
ShanghaiTech University
} 

\maketitle

\begin{abstract}
	

The log-structured merge tree (LSM-tree) 
gains wide popularity in
building key-value (KV) stores. 
It employs logs to back up arriving KV pairs and maintains
a few on-disk levels with exponentially increasing capacity limits,
resembling a tiered tree-like
structure. 
A level comprises 
SST files, each of which 
holds a sequence of sorted KV pairs.
From time to time,
LSM-tree redeploys KV pairs from a full level to the lower level
by compaction, which merge-sorts and moves
KV pairs among SST files, thereby incurring substantial disk I/Os.

In this paper, we revisit the design of LSM-tree and find
that organizing multiple KV pairs in an SST file entails the heavyweight redeployment
of actual KV pairs in a compaction. 
Accordingly we revolutionize the organization of KV pairs
by transforming an SST file of KV pairs to an {\em SST directory}, in which each KV pair 
makes into an independent {\em KV file}
with the key and value as filename and main file contents, 
respectively. 
Moving KV pairs in a compaction
converts to transferring directory entries ({\tt dentry}s), which causes concretely fewer disk I/Os.
This is the essence of our design named {\bf \DeLSM}. 
We build a prototype of \DeLSM on LevelDB and evaluation results show
that it significantly outperforms the state-of-the-art LSM-tree variants in different dimensions.

\end{abstract}

\section{Introduction}\label{sec:intro}

The log-structured merge tree (LSM-tree) 
plays an important role in
building key-value (KV) stores
to service omnipresent workloads, such as web indexing, social networking, and e-commerce
~\cite{LSM:bigtable:OSDI-2006,LSM:LevelDB,LSM:RocksDB,LSM:HBase,LSM:Cassandra:2010,LSM:NoveLSM:ATC-2018,LSM:SLM-DB:NVM:FAST-2019}.
LSM-tree is a tiered tree-like structure composed of multiple levels
in memory and on disk. 
It employs in-memory {\em memtables} to absorb  
arriving 
KV pairs after appending them to an on-disk log for crash consistency. 
A shortage of memtable space triggers a {\em minor compaction} that
dumps a full memtable into 
a {\em sorted string table} (SST) file to be placed 
at the top level (L$_0$) on disk.
Each on-disk level 
has a fixed capacity limit and L$_n$ 
is typically 10$\times$ smaller than L$_{n+1}$ ($n\geq0$).
When L$_n$ becomes full, LSM-tree initiates a {\em major compaction} which 
 merge-sorts KV pairs from selected SST files residing in L$_n$ and L$_{n+1}$.
It then orderly moves and packs involved KV pairs into new L$_{n+1}$ SST files.

A major compaction results in substantial disk I/Os and impairs performance
due to redeploying KV pairs among SST files. Worse,
it may stall LSM-tree from serving foreground requests.
Moreover, a KV pair would be factually written  
again whenever
it is involved in a further compaction, causing severe write amplification.
Researchers 
have looked into how to reduce the performance penalty of LSM-tree~\cite{LSM:LSM-trie:ATC-2015,LSM:WiscKey:FAST-2016,LSM:TRIAD:ATC-2017,LSM:LSbM-tree:ICDCS-2017,LSM:LWC-tree:TOS-2017,LSM:PebblesDB:SOSP-2017,LSM:SlimDB:VLDB-2017,LSM:FlashKV-OpenSSD:TECS-2017,LSM:NoveLSM:ATC-2018,LSM:LDS:TPDS-2018,LSM:Accordion:VLDB-2018,LSM:LDC-SSD:ICDE-2019,LSM:FPGA-accelerated-compaction:FAST-2020,LSM:FPGA-based-compaction:ICDE-2020,LSM:GearDB:ATC-2020,LSM:BoLT:Middleware-2020,LSM:L2SM:ICDE-2021,LSM:Nova-LSM:SIGMOD-2021}. Some of them focused on emerging storage devices~\cite{LSM:WiscKey:FAST-2016,LSM:FlashKV-OpenSSD:TECS-2017, LSM:NoveLSM:ATC-2018,LSM:LDC-SSD:ICDE-2019,LSM:SLM-DB:NVM:FAST-2019,LSM:GearDB:ATC-2020}, while others sought the aid from specific hardware~\cite{LSM:FPGA-accelerated-compaction:FAST-2020,LSM:FPGA-based-compaction:ICDE-2020}.  
In this paper, however, we  
envisage a solution to revolutionize LSM-tree, which shall be 1) simplistic but efficient, and 2)
generic and software-only, 
for the ease of development and portability.

With this intention, we have studied LevelDB, one typical LSM-tree-based KV store,
with emphasis on its data structures and compaction procedure.
Through the intensive study, we have obtained several key observations.
\begin{itemize}[leftmargin=4mm]\setlength{\itemsep}{-\itemsep}	
	\vspace*{-0.7ex}
	\item An SST file indexes multiple KV pairs stored in it, which shares similarities with a directory of file system indexing files with directory entries ({\tt dentry}s).  
	Although users put individual KV pairs in the foreground,  
	LevelDB treats an SST file as the unit for processing in the background.
	\item SST files are the core of compaction.
	A compaction is a series of unpacking, sorting, and redeploying KV pairs with a few 
	SST files through disk reads and writes.
	\item An SST file remains immutable (read-only) in its lifetime, i.e., since its creation until it is compacted. In fact, once put into LevelDB, a KV pair stays unchanged
	unless it subsequently encounters a future deletion.
	\item LevelDB relies on the underlying file system to manage SST files. To get a KV
	pair, LevelDB waits for file system to fetch the {\tt dentry} of 
	corresponding SST file and then searches in the file.
	In each compaction, file system adds {\tt dentry}s for LevelDB
	to index new SST files. Such 
	 is much more lightweight  
	than moving KV pairs within these files.
	\vspace*{-0.7ex}	
\end{itemize}

Inspired by these observations, we consider 
reorganizing KV pairs for LSM-tree by leveraging the fundamental idiosyncrasies of 
file system, i.e., the directory-files structure. 
In a nutshell, we transform an SST file to an {\em SST directory}
and make each KV pair into an individual {\em KV file}. 
A compaction converts to a succession of efficient {\tt dentry} operations between 
 SST directories. 
This is the essence of our design, namely \DeLSM ({\tt \underline{De}ntry}-based \underline{LSM}-tree).
The main ideas of \DeLSM  
 are summarized as follows.
\begin{itemize}[leftmargin=4mm]\setlength{\itemsep}{-\itemsep}	
	\vspace*{-0.7ex}
	\item \DeLSM processes a KV pair as the integral unit in a compaction. 
    In a minor compaction dumping a memtable, it
	 creates a KV file for each KV pair,   
	with the key and value being the filename and main file content, respectively.
	\DeLSM retains the mechanism of logging and memtables like standard LSM-trees but
	schedules a minor compaction to be performed in the background
	once a memtable becomes immutable.	
	\item \DeLSM manages a {\em level directory} to hold SST directories of each level.
	A major compaction between L$_n$ and L$_{n+1}$ ($n\geq0$) directories is a succession of
	linking and unlinking KV files ({\tt dentry}s) 
	over existing and new SST directories, thereby incurring drastically
	fewer disk I/Os compared to conventional redeployment of actual KV pairs.
	\item \DeLSM basically serves a search request with the underlying file system 
	traversing {\tt dentry}s to locate a key or key range (point or range search).
	For optimization, it employs a local meta file with SST-level Bloom
	filter in each SST directory and global structures that keep key ranges of all
	valid SST directories.
	\vspace*{-0.7ex}
\end{itemize}

The avoidance of  
moving KV pairs for compactions 
enables \DeLSM's high efficacy. 
\DeLSM also embraces high portability as 
a software-only solution with neither hardware tailoring
or customization nor any onerous change of file system, 
We have implemented a prototype for it with
LevelDB and Ext4. Evaluation results show that
\DeLSM boosts the performance of LevelDB up to 18.9$\times$ and 2.8$\times$, respectively,
for write and read requests. It also significantly outperforms 
LSM-tree variants including 
WiscKey~\cite{LSM:WiscKey:FAST-2016}, PebblesDB~\cite{LSM:PebblesDB:SOSP-2017},
LDS~\cite{LSM:LDS:TPDS-2018}, and BoLT~\cite{LSM:BoLT:Middleware-2020} in different dimensions.

The remainder of this paper is organized as follows.
In Section~\ref{sec:background} we briefly introduce LSM-tree with LevelDB.
In Section~\ref{sec:motivation} we analyze observations obtained
in the aforementioned study. We detail the design and implementation of
\DeLSM in Section~\ref{sec:design}.  
We describe the evaluation of \DeLSM in Section~\ref{sec:evaluation}.
In Section~\ref{sec:related}, we discuss  
relevant LSM-tree variants. We conclude the paper in Section~\ref{sec:conclusion}.

\begin{figure}[t]
	\centering
	\includegraphics[width=\columnwidth]{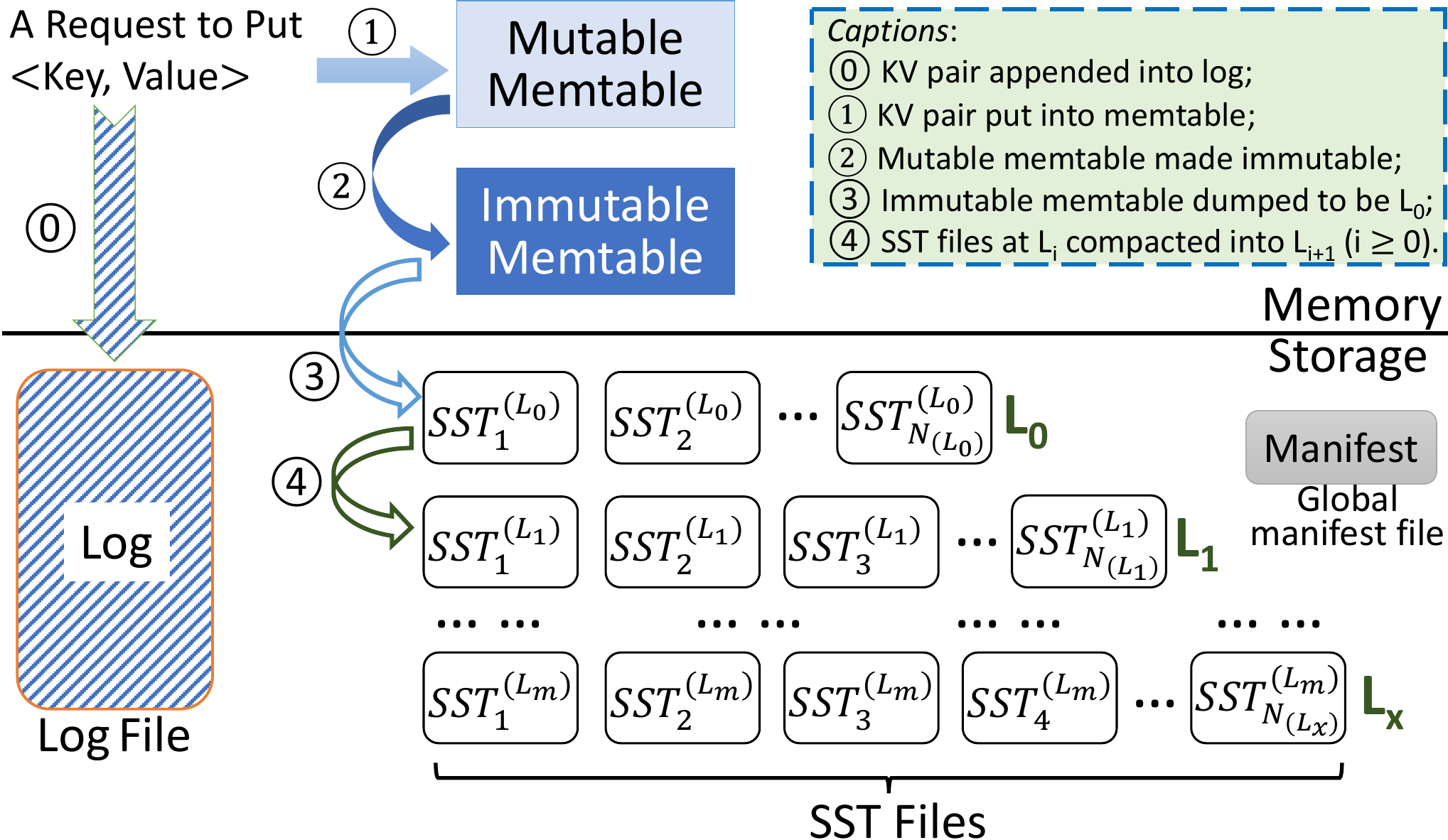}
	\caption{LevelDB's Main Components}\label{fig:LSM}
	\vspace*{-2ex}	
\end{figure}

\section{Background}\label{sec:background}
\subsection{LSM-tree}\label{sec:bg:LSM}

LevelDB~\cite{LSM:LevelDB} is a typical KV store based on LSM-tree.
Without loss of generality, we 
take LevelDB to illustrate the structure of LSM-tree and
how it serves write and read
requests.

\autoref{fig:LSM} sketches the architecture of LevelDB.
It leverages on-disk logs for crash consistency and
maintains memtables in memory and mutiple levels on disk.
Each level is a sorted run of many KV pairs stored in SST files.
The capacity limit of
a level is at least ten times larger than the level immediately above it,
thereby resembling a hierarchical tree-like structure.

\textbf{Write request and compaction}\hspace{1ex}
On a write request (e.g., {\tt Put}),
LevelDB appends the KV pair to the log
(\circled{0} in~\autoref{fig:LSM}). After that,
LevelDB inserts it into the mutable memtable (\circled{1}).
These are two main steps happening in the foreground of LevelDB.
A full memtable is made
immutable and 
 a new mutable one 
is created for incoming KV pairs (\circled{2}).

When the space of mutable memtable is insufficient,
LevelDB triggers a {\em minor compaction} (\circled{3}). It
creates an SST file at L$_0$ 
and dumps KV pairs of the immutable memtable
to the file.
Since L$_{0}$ SST files are directly converted from memtables
that have been receiving KV pairs from users,  
key ranges may overlap between them.
When the size of all L$_{0}$ SST files reaches L$_{0}$'s capacity 
limit, LevelDB merges KV pairs from L$_0$  
and selected L$_{1}$ SST files into new L$_{1}$ SST files 
that rule out overlapping key ranges.
This process is called {\em major compaction} (\circled{4}) that can
occur  
between
any L$_n$ and L$_{n+1}$ ($n\geq 0$),   
as every level has a capacity limit. 
Compactions badly impair performance due to sorting, moving, and persisting
KV pairs with on-disk files.  
Worse, if the mutable memtable is full but a minor compaction
is still ongoing,  
LevelDB will  
stall serving requests in the foreground
until the compaction is completed.

\textbf{Read request}\hspace{1ex}
Upon a {\tt Get} request to read a KV pair (point search), 
LevelDB successively searches mutable memtable, immutable memtable,
L$_{0}$, L$_{1}$, and so on, since a newer version for the key
 exists in a higher level.
As to a range search that demands a number of KV pairs, e.g.,
keys starting with `a', the organization of
contiguously sorted keys in the memtables
and SST files is supportive. Yet keys in a 
range may reside in multiple
SST files, possibly across several levels~\cite{LSM:REMIX-range-search:FAST-2021}.

\textbf{Log}\hspace{1ex}
LevelDB dedicates a log file to a memtable.
It appends a KV pair with operation 
(e.g., {\tt Put}) to the tail of  
log file before an insertion with the mutable memtable.
LevelDB discards a log file after 
compacting its corresponding memtable, 
as the latter has been transformed into an  
 L$_0$ SST file.

\begin{figure}[t] 
	\centering
	\includegraphics[width=\columnwidth]{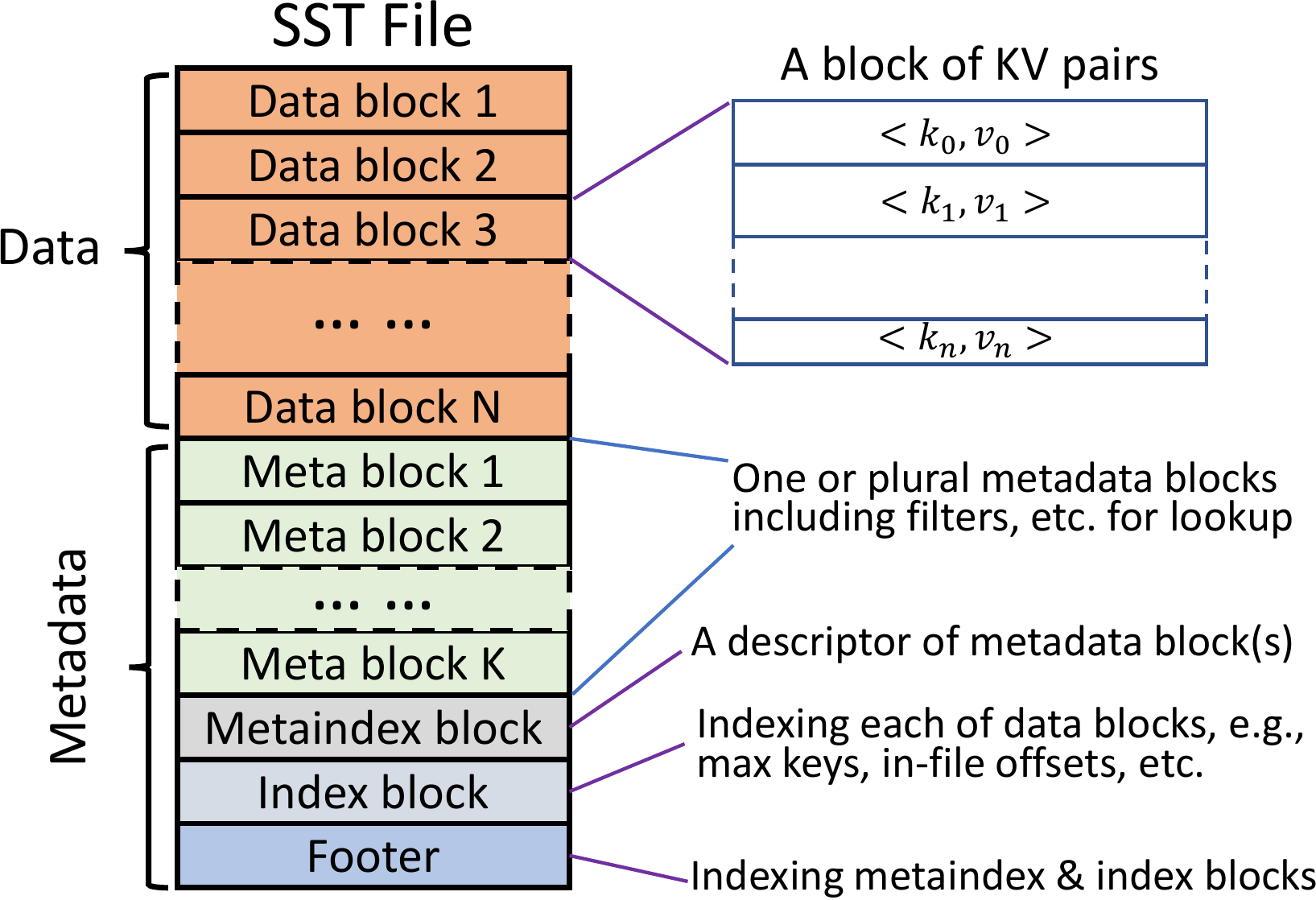}
	\caption{LevelDB's SST File}\label{fig:SST}
	\vspace*{-2ex}
\end{figure}

\textbf{SST}\hspace{1ex} 
\autoref{fig:SST} shows the format of an SST file.
It begins with multiple data blocks comprised of sorted KV pairs.
Next there are meta blocks  
used to index KV pairs in data blocks.
A meta block mainly contains a sequence of block-level Bloom Filters
that can help LevelDB quickly determine if a key is in the SST file
or not.  
The subsequent metaindex and index blocks record in-file offsets that 
track meta and data blocks, respectively.
The last footer field  
holds in-file offsets to the metaindex and index blocks.
\autoref{fig:SST} summarizes the functions of these fields.
An SST file at any level 
is immutable ({\em read-only}), and remains alive until it is 
compacted into the next level.
LevelDB also employs a global manifest file to track
valid SST files with their respective key ranges.

\subsection{Directory Management of File System}\label{sec:bg:dentry}

LevelDB defines rules to name its SST, log, and manifest files.
It puts them in directories and relies on the underlying file system
to manage them. 
File system makes each directory into 
a specific file comprised of {\em directory entries}
({\tt dentry}s).
A {\tt dentry} records the mapping from a filename to the file's
{\tt inode} number, with which the file's contents can be retrieved.

Using {\tt dentry}s to organize directory and files 
is one of the fundamental idiosyncrasies
of file system. A directory of {\tt dentry}s and 
all files in it are all permanently stored on disk
and file system must efficiently index them. For example,
Ext4 employs a hash tree (HTree) to 
do so. It hashes each filename and uses the hash value
as a key to build a B-tree over a directory's {\tt dentry}s.  
Given a filename, Ext4 loads the HTree into memory and
searches it to locate the file
by calling {\tt ext4\_lookup}.
To fetch multiple files with a pattern,
e.g., any file with a name starting with `a',
Ext4 calls {\tt ext4\_readdir} that 1)
builds an in-memory red-black tree (RB-tree) with {\tt dentry}s
loaded from the HTree,
and 2) traverses the RB-tree to filter out files that match
the pattern.
Obviously, to get one or a few files in
a large number of {\tt dentry}s is not efficient
for either B-tree or RB-tree.
The operating system hence employs a global {\em {\tt dentry} 
cache} ({\tt dcache}) to buffer {\tt dentry}s
that are recently and frequently used.
Note that LevelDB also maintains a table cache in which
the file descriptors of SST files are buffered for quick
reference~\cite{LSM:LevelDB, LSM:BoLT:Middleware-2020}.

\section{Motivation}\label{sec:motivation}

By running LevelDB on top of Ext4, we have conducted a 
 study, especially on the similarity and interaction between them.
All the disk I/Os that LevelDB incurs for heavyweight compactions 
are performed by Ext4. 
Our aim with the study is to find a way that makes the most of file system
to minimize disk I/Os for LSM-tree.

\textbf{\textgoth{O}1. LevelDB receives individual KV pairs issued by users 
in the foreground but packs them into a large SST file as the 
unit for subsequent processing in the background.}

KV pairs are individually put by users. 
To take advantage of sequential writes of storage devices,
LevelDB appends them to a log in the foreground.
However, LevelDB bundles multiple KV pairs into an on-disk SST file that is used
 as the unit for background processing.
An SST file 
is generally configured to have a large size, e.g., 4MB.
Assuming that a KV pair takes 512B on average, 
about 8,000 KV pairs stay in one SST file.
It is non-trivial to frequently merge-sort and redistribute
so many KV pairs from multiple SST files during a compaction.

\textbf{\textgoth{O}2.  
	The way LevelDB organizes and indexes KV pairs in one SST file shares similarities with file system's {\tt dentry} management strategy.}

To trace and index data is a fundamental functionality for any application- or system-level data management system.
LevelDB encapsulates metadata in an SST file to index stored KV pairs, while
Ext4 employs an HTree with {\tt dentry}s to index SST files 
in accordance with  their filenames. 
To move or read a KV pair needs to
go through both levels of indexes.

\textbf{\textgoth{O}3. A compaction that moves KV pairs between SST files is severely inefficient 
	compared to file system moving files ({\tt dentry}s) between directories.}

In a compaction, LevelDB redeploys KV pairs from existing SST files
into newly-created ones, resulting in substantial disk read and write I/Os.
Comparatively, when Ext4 moves files between directories, it just moves  
{\tt dentry}s, which causes insignificant I/Os as the size of a {\tt dentry} is generally small.

\textbf{\textgoth{O}4. Most on-disk KV pairs stay alive and unchanged for a long time
	but they would migrate from one SST file to the next one due to compactions.} 

The life of an SST file ceases once KV pairs it holds have been compacted.
A KV pair yet remains alive and unchanged on disk since dumped into L$_0$, 
unless it encounters a deletion propagated from upper levels.
Whereas, considering the compaction algorithm of LevelDB,
a KV pair does not stay at a specific on-disk location but
is repeatedly transferred by Ext4 from one
SST file to the other one at the lower level
once a further compaction happens to it.

In light of foregoing observations, we conclude that
1) to reorganize an individual KV pair as the independent unit for processing aligns with 
the nature of KV store and promises flexibility,
and 2) the similarity
between an SST file with KV pairs and a directory with 
{\tt dentry}s implies a potential collaboration of file system
and LSM-tree to  
revolutionize the latter's organization of KV pairs and in turn
gain higher performance
with fewer disk I/Os.

\section{Design of \DeLSM}~\label{sec:design}

As the conventional organization of KV pairs
causes heavyweight compactions, we 
propose \DeLSM that leverages
file system, particularly the directory-files structure, 
to reorganize KV pairs for LSM-tree.
\autoref{fig:arch} shows the architecture of \DeLSM. 
Firstly, 
\DeLSM 
reformats an SST file
into an {\em SST directory}, in which
each KV pair exists as an independent {\em KV file} with a {\tt dentry} (Section~\ref{sec:SSTD}).
Secondly, 
during a compaction, \DeLSM moves {\tt dentry}s for KV files 
instead of redeploying actual KV pairs between SST directories, 
thereby concretely improving the efficiency of compactions 
(Section~\ref{sec:compact}).
Thirdly, besides utilizing file system's native {\tt dentry} management 
for  
searches,
\DeLSM employs a {\em meta file} with an SST-level Bloom filter in each SST directory  
(Section~\ref{sec:search}).
\DeLSM also guarantees the crash consistency of stored KV 
pairs~(Section~\ref{sec:consistency}).
Last but not the least, \DeLSM is effectually implementable 
with a generic file system~(Section~\ref{sec:implementation}).

\subsection{KV File and SST Directory}\label{sec:SSTD}

\DeLSM reorganizes a KV pair as an individual KV file. 
It appoints the key and value as the filename and main file
content, respectively, since the two can be viewed as byte-arrays (strings).
In each KV file, there is a header that 
contains
the operation code for the KV pair, 
e.g., `0' for {\tt Put} and `1' for {\tt Delete},
and a globally monotonic increasing sequence number.
\DeLSM groups multiple KV files in 
an SST directory,
the name of which follows LevelDB's naming rule
for SST files. 
Consequently, \DeLSM delegates file system to manage  
KV files through {\tt dentry}s  
in an SST directory, which obviates the metadata LevelDB puts down
in each SST file (cf.~\autoref{fig:SST}). 
\DeLSM adds a supplemental {\em meta file} in an SST 
directory for faster lookup (cf. Section~\ref{sec:search}).
In addition, \DeLSM makes a {\em level directory} 
for each level, i.e., L$_{n}$ directory ($n\geq0$),
in which it accommodates all SST directories belonging to L$_{n}$.

\begin{figure}[t]
	\centering
	\includegraphics[width=\columnwidth]{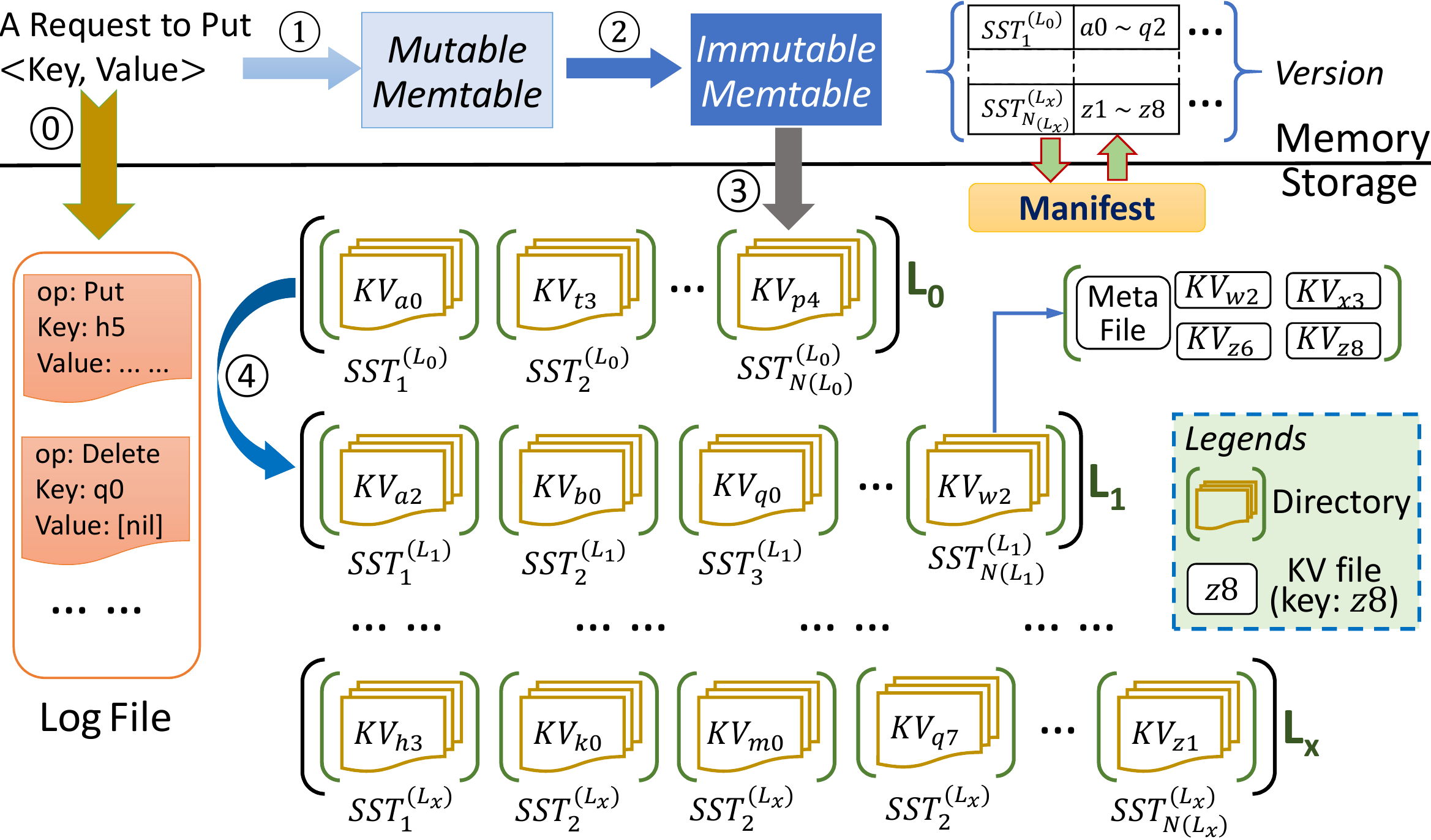}
	\caption{The Architecture of \DeLSM}\label{fig:arch}
\end{figure}

An SST directory corresponds to a memtable.
In the foreground, before inserting KV pairs into the mutable memtable,
\DeLSM still uses a log to absorb them, as logging generates sequential
writes favored by storage devices.
When the mutable memtable is full, \DeLSM marks it to be immutable 
and creates a new mutable one. In the meantime, \DeLSM starts
a minor compaction in the background 
and transforms each KV pair in the immutable memtable to be a KV file.
This can be done through a background thread (cf.~Section~\ref{sec:implementation}).

As shown in the top-right of~\autoref{fig:arch}, like LevelDB,
\DeLSM maintains a global manifest file that backs an
in-memory structure known as the {\em version} 
to track all valid SST directories. 
In brief, the version
1) monitors the increase of sequence number, 
2) regulates
the naming of logs and SST directories,   
3) records
the addition and removal of SST directories for a compaction,
and so on.

\subsection{Lightweight Compaction}\label{sec:compact}

KV files remain in SST directories until selected for a major compaction,
in which
\DeLSM moves their {\tt dentry}s
rather than themselves.
As a result,
\DeLSM redefines the capacity limit 
of a level as the number of {\tt dentry}s
for KV files, instead of the size of KV pairs.
The capacity limit of a level is still ten times that of the upper one for \DeLSM.

When L$_n$ ($n \geq 0$) reaches its capacity limit, \DeLSM initiates
a major compaction (\circled{4} in~\autoref{fig:arch}). 
In existing L$_n$ and L$_{n+1}$ SST directories,
\DeLSM finds out appropriate keys by iterating
over {\tt dentry}s and inserts them
into a newly-created L$_{n+1}$ SST directory.
When the new L$_{n+1}$ SST directory becomes full, 
\DeLSM creates the next one and continues moving {\tt dentry}s 
until all involved keys are compacted.
Then it removes {\tt dentry}s of compacted SST directories from L$_n$ and L$_{n+1}$.
At the end of a compaction, \DeLSM records the change of SST directories 
in the in-memory version and manifest file. The key ranges of newly-generated SST directories
are also added to the version for ease of searches (cf. Section~\ref{sec:search}).
As \DeLSM's major compaction is mainly composed of 
finding and redeploying {\tt dentry}s rather than 
moving KV pairs, 
it is more lightweight and efficient with much fewer disk I/Os. 

\begin{figure}[t] 
	\centering
	\includegraphics[width=\columnwidth]{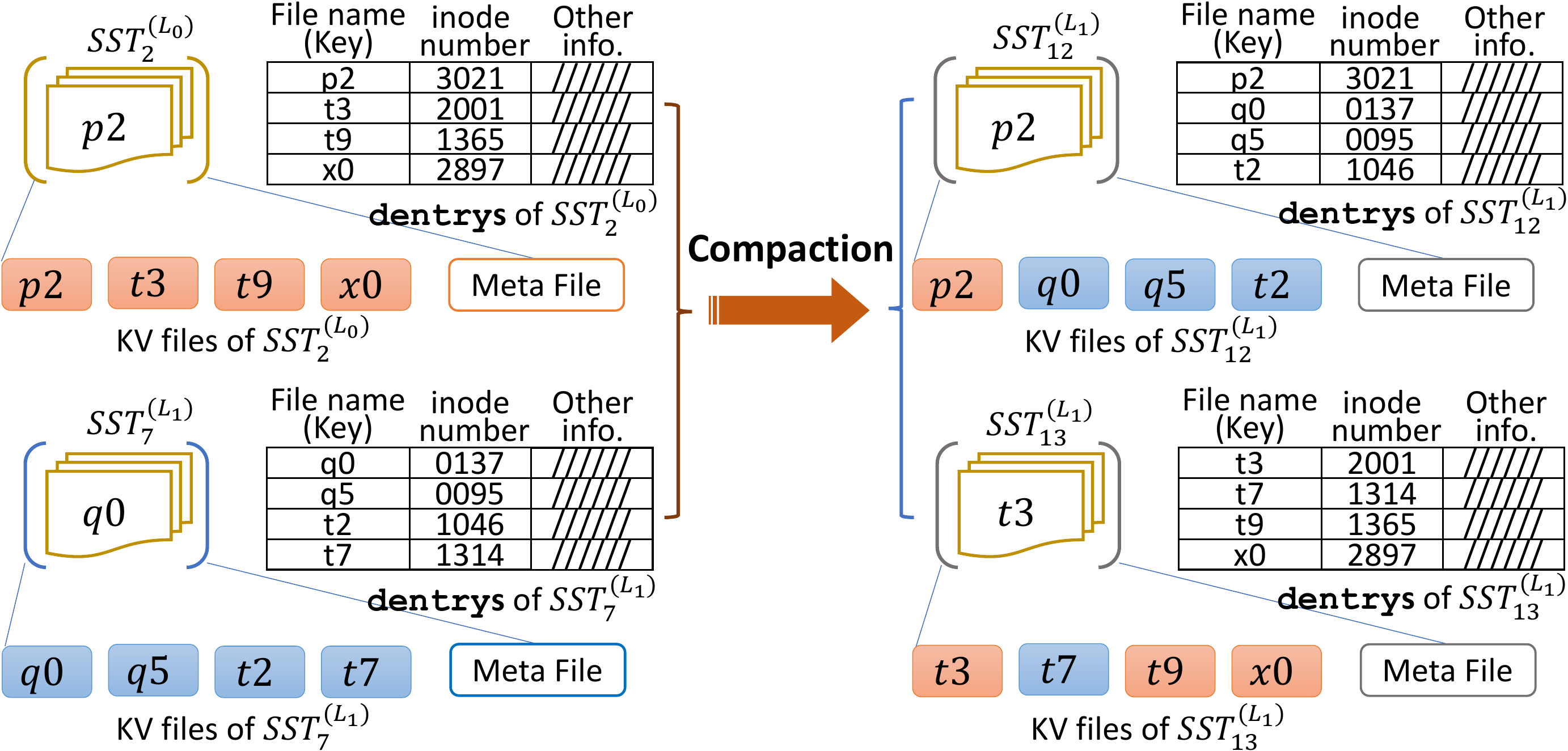}
	\caption{A Compaction of \DeLSM}\label{fig:compact}
\end{figure}

A deletion or overwrite of a KV pair is also conducted in the process of 
compaction. When \DeLSM encounters two files with the same filename,
it handles them regarding their operation codes and sequence numbers.
The KV file with a greater sequence number decides
the KV pair's fate. For example, given one KV file with
a deletion code and a greater sequence number, \DeLSM 
shall remove the other one.

\autoref{fig:compact} illustrates an example of 
compacting two SST directories ($SST_2^{(L_0)}$ and $SST_7^{(L_1)}$)
into L$_1$. Before the compaction, each SST directory contains keys in different ranges.
The compaction creates new L$_1$ directories ($SST_{12}^{(L_1)}$ and $SST_{13}^{(L_1)}$),
and redeploys {\tt dentry}s into them. 
As a compaction of \DeLSM is mainly a series of directory operations, 
a comparison between {\tt dentry}s before and after the compaction in \autoref{fig:compact}
indicates that KV files and their {\tt inode} numbers remain the same as \DeLSM incurs no change to them.

\subsection{Point Search and Range Search}\label{sec:search}

On a point search for a key,
\DeLSM first checks memtables, which hold
the latest updates to KV pairs.
A miss leads \DeLSM to L$_0$.
L$_0$ SST directories contain overlapping key ranges.
\DeLSM asks the version  
for all L$_0$ SST directories that the target key falls in.
In a candidate SST directory, \DeLSM needs file system
to find if the key (filename) exists among {\tt dentry}s.
A match returns the target value,
i.e., file content pointed by the {\tt dentry}'s {\tt inode}.
Otherwise, \DeLSM does with the next candidate.
If the key is not found in L$_0$,
\DeLSM goes down to L$_1$ that is with non-overlapping key ranges
and searches in one candidate SST directory. 
It continues until the key is found
or traverses all levels without a hit.
Given a range search to get, 
say, keys starting with `a', 
\DeLSM still goes through memtables and on-disk levels.
It obtains SST directories with appropriate key ranges from the version 
and traverse {\tt dentry}s for KV files accordingly.  

Relying on file system to search keys through {\tt dentry}s
is sound but not very efficient.  
Take Ext4 for instance. To get one KV file for point search, Ext4
needs to first load on-disk nodes of an HTree (B-tree) into memory.
Regarding L$_0$ SST directories with overlapping key ranges,
searching the key in non-target SST directories incurs
substantial disk I/Os. To reduce such performance penalties,
\DeLSM encapsulates a {\em meta file} in each SST directory.
It makes and puts an SST-level Bloom filter for all keys 
in the meta file when the SST directory is being built during 
a compaction. Consequently, if a
key is not hit in the Bloom filter,
\DeLSM no longer scans {\tt dentry}s of an SST directory.
Note that \DeLSM's SST-level Bloom filter differs 
from the block-level Bloom filters used by LevelDB which stores KV pairs contiguously in disk blocks of an SST file~\cite{LSM:LevelDB, LSM:PebblesDB:SOSP-2017}.
Also, in contrast to LSM-trees
that hold numerous metadata in an SST file for indexing (cf.~\autoref{fig:SST}),
\DeLSM only keeps one Bloom filter, since file system
helps to index KV pairs for \DeLSM. This both gains
space efficiency and simplifies management.

Besides the meta file, \DeLSM also maintains a block cache
and a table to buffer frequently used 
KV pairs and file descriptors of KV files~\cite{LSM:LevelDB,LSM:HBase,LSM:LSbM-tree:ICDCS-2017,LSM:AC-key-cache:FAST-2020}, respectively.
These caches further accelerate search operations for \DeLSM.
In the future, we would explore how to jointly manage and
utilize \DeLSM's user-space caches with kernel-space buffers, such as
the page cache, {\tt dentry} cache and {\tt inode} cache.

\subsection{Crash Consistency of \DeLSM}\label{sec:consistency}

Conventional LSM-trees 
log KV pairs before putting them in memtables.  
For KV pairs involved in a minor or major compaction,  
LSM-trees sync new SST files by {\tt fdatasync} or {\tt fsync}
before removing obsolete logs or compacted SST files.  
Comparatively,
\DeLSM also utilizes logging to guarantee 
the consistency of arriving KV pairs to be put in memtables. 
The minor compaction is the only occasion on which
\DeLSM generates KV files from an immutable memtable.
\DeLSM does not sync KV files though, as syncing multiple small
files is costly. 
Instead, \DeLSM syncs a full log file 
when a memtable is made immutable. 
Next
\DeLSM transiently retains the synced log file 
until an approximate duration, say, 60 to 120 seconds, elapses
after the generation of KV files, in order 
to secure the consistency and durability of generated KV files.
This duration gives file systems like Ext4 
sufficient time to commit data and metadata of KV files into 
disk~\cite{FS:optimistic-consistency:SOSP-2013, FS:iJournaling:ATC-2017}.

On the other hand, like LevelDB and other LSM-tree variants,
\DeLSM leverages the manifest file for recovery.
Either a major or minor compaction of \DeLSM involves 
the addition and/or removal of  
SST directories. 
\DeLSM composes, appends, and syncs  
a record of the changes of SST directories to the manifest file.
As a result, the manifest file contains all traceable changes over time.
 
A record synced to the manifest file marks the end of a compaction
and enables the recoverability of \DeLSM. 
In case of a crash, \DeLSM loads and checks
the manifest file to rebuild the in-memory version that tracks valid  
SST directories. Upon 
a corrupted record of 
adding and removing SST directories, 
\DeLSM does with relevant directories
 and makes a new proper record.  
No corrupted record does not rule out  
inconsistency as the crash might have happened before
writing the record or during a compaction.
\DeLSM scans the manifest file, log files, and directories across  
all levels. Each KV pair is stored along with a monotonic increasing
sequence number in a log or an SST directory while
the manifest file has the last valid sequence number.
\DeLSM is thus able to identify all legally valid
KV pairs for proceeding to a consistent state.

\subsection{Implementation}\label{sec:implementation}

\begin{table}
	\resizebox{\linewidth}{!}{
		\begin{tabular}{l | l|l}
			\toprule
			\multicolumn{2}{l|}{\DeLSM functions} & File system's supports with system calls \\ \hline \hline	
			\multicolumn{2}{l|}{\multirow{2}{*}{Put a KV pair (file)}} & {\tt open} to create the KV file, and {\tt write} to \\
			\multicolumn{2}{l|}{}& store the value into the file \\ \hline
			\multicolumn{2}{l|}{Delete a KV pair (file)} & \multirow{2}{*}{{\tt remove} to remove a file or directory} \\ \cline{1-2}
			\multicolumn{2}{l|}{Remove an SST directory}& \\ \hline
			\multicolumn{2}{l|}{Create a directory}& {\tt mkdir} to create a new SST directory \\	\hline		
			\multicolumn{2}{l|}{\multirow{3}{*}{Major Compaction}} & {\tt mkdir} to create new SST directories,  {\tt link}  \\
			\multicolumn{2}{l|}{}&to redeploy the {\tt dentry}s by hard links, and \\
			\multicolumn{2}{l|}{}&  {\tt remove} to remove existing SST directories  \\ \hline	
			\multirow{2}{*}{Search}& Point & {\tt lookup} to find a {\tt dentry} in SST directory\\ \cline{2-3}
			&Range & {\tt readdir} to check {\tt dentry}s in SST directory \\ 
			\bottomrule
	\end{tabular}}
	\caption{Underlying Supports of File System for \DeLSM}\label{tab:impl}
\end{table}

\textbf{The supports of file system}\hspace{2ex} 
As shown in~\autoref{tab:impl},
all the supports demanded by \DeLSM from file system can be
satisfied by standard POSIX-compliant system calls.  
Putting a new KV pair is creating a file with the key as filename
via {\tt open} and writing down the value 
via {\tt write}. Deleting an on-disk KV pair is removing the KV file and {\tt dentry} 
via {\tt remove}.
For compactions, \DeLSM mainly makes use of {\tt mkdir} to create 
directories and {\tt link} to redeploy {\tt dentry}s.
There are two types of links, i.e., hard
or symbolic link. The hard link factually refers
to the {\tt inode} of a file.
After creating new SST directories, \DeLSM sets hard links to
KV files to be moved.
At last, \DeLSM removes compacted {\tt dentry}s, files, and SST directories
by successively calling {\tt remove}.  
To sum up, the implementation of \DeLSM is viable with a 
generic file system.

\textbf{Compactions} \hspace{2ex}
Considering the speed mismatch between putting KV pairs into memtable 
in the foreground and generating a KV file in the background, 
\DeLSM transiently keeps multiple immutable memtables to be 
compacted,  
instead of stalling 
to wait for the completion of a minor compaction.
It employs a background thread to handle those memtables.
On the other hand, \DeLSM triggers a major compaction 
if the number of KV files ({\tt dentry}s) at a level reaches
a limit that is configurable. For example, we set the limit
for L$_0$ as 10,000 after empirical tests with our machine.

\textbf{The limits of filename}\hspace{2ex}
Using the key as filename entails three issues. 
One is the maximum length of a filename. Most file systems
allow up to 255 bytes for a filename. With 
regard to state-of-the-art studies, keys are generally in
dozens of bytes~\cite{LSM:Facebook:FAST-2020}. Therefore, the limit of 
255 bytes is not a thwarting concern.
Moreover, such a length limit is adjustable by slightly changing  
the implementation of file system.

The second issue is that,  
despite supporting tens of thousands of diverse characters
from various human languages,  
a filename fundamentally denies a handful of characters.
For example, Linux forbids  
the forward slash (`/') 
in a filename.
When \DeLSM receives a key with such illegal characters, 
a straightforward solution is to return the key back 
with a revision suggestion.
This is empirically feasible and used in our implementation. 
In practice, many online service providers, like Google and Microsoft,
define respective rules to refuse a few characters in their account names, such as forward slash (`/')
 and asterisk (`*').

Thirdly, file system does not allow files with the same filename to
exist in one directory. However, standard LSM-trees
provide a snapshot feature that retains all variants
of a KV pair with sequence numbers no smaller than the sequence number
at which the snapshot option has been switched on.
To support snapshots, \DeLSM
puts values  
with the same key into one KV file 
according to the ascending order of their sequence numbers. 
Hence there are no KV files with the same 
filename in an SST directory.
If a user aims to access a key's value for 
a specific snapshot saved at some sequence number,
\DeLSM retrieves from the KV file variants with 
sequence numbers no smaller than the snapshot's.

\textbf{The impact of {\tt sync} and system calls}\hspace{2ex}
\DeLSM calls {\tt sync} to persist logs 
and {\tt dentry}s for consistency and durability. Although \DeLSM
avoids repeatedly syncing KV pairs in compactions, the impact of {\tt sync} still exists~\cite{FS:iJournaling:ATC-2017, LSM:BoLT:Middleware-2020}. Recently Ext4 has included a promising feature called {\em fast commit} with improved {\tt sync}~\cite{FS:fast-commit,FS:iJournaling:ATC-2017}.
We leave a configurable option for users to turn on or 
off {\tt sync}. 
In addition, LevelDB was reported to exhibit errors upon the failure of {\tt sync}~\cite{FS:Fsync-fails:ATC-2020}. Any system call has a likelihood
of failure. We attend such failures comprehensively within the implementation
of \DeLSM.

The next concern is that \DeLSM
operates more files and {\tt dentry}s through system calls than standard LSM-trees
and also transfigures many sequential I/Os to be random ones. 
The overhead of file system calls and their impact in the context of LSM-trees have been investigated~\cite{FS:Moneta-D:ASPLOS-2012,FS:SplitFS:SOSP-2019,FS:CrossFS:OSDI-2020,LSM:LDS:TPDS-2018,OS:KLOCs:ASPLOS-2021}. There is a trade-off between moving actual KV pairs
and handling more system calls with random I/Os.
Our evaluation in Section~\ref{sec:evaluation} 
confirms that, due to the overwhelming disk I/Os caused by conventional compactions, 
\DeLSM's reorganization 
of KV pairs in the way of directory-files is worthwhile and efficient.

\section{Evaluation}~\label{sec:evaluation}

We evaluate \DeLSM against state-of-the-art LSM-tree variants
with micro-benchmarks for preliminary tests. We mainly aim to
answer the following questions.
\begin{itemize}[leftmargin=4mm]\setlength{\itemsep}{-\itemsep}	
	\vspace*{-0.7ex}
	\item Does \DeLSM's avoidance of moving actual KV pairs yield high write performance?
	\item How significant is the drop of I/O amplification caused by \DeLSM?
	\item How is the read performance of \DeLSM?
	\vspace*{-0.7ex}	
\end{itemize}

\subsection{Evaluation Setup}

\textbf{Platform}\hspace{1ex} We have run all experiments on a server with 1) 64-core Intel Xeon Gold 5218 CPU, 2) 192GB memory,
3) 1TB SSD, and 4) Ubuntu 20.04.2 with Linux 5.11.0 and GCC/G++ 9.3.0. In particular, the
underlying file system is Ext4 mounted in the default `data=ordered' mode.

\textbf{Competitors}\hspace{1ex}
Besides the original LevelDB, we considered state-of-the-art LSM-tree variants including 
WiscKey~\cite{LSM:WiscKey:FAST-2016}, LDS~\cite{LSM:LDS:TPDS-2018}, PebblesDB~\cite{LSM:PebblesDB:SOSP-2017},
and BoLT~\cite{LSM:BoLT:Middleware-2020}.  
The latter two have official source codes\footnote{BoLT: \url{https://github.com/DICL/BoLT}}\footnote{PebblesDB: \url{https://github.com/utsaslab/pebblesdb}} while we implemented the former two and our \DeLSM atop LevelDB in accordance with their respective designs. 
These LSM-tree variants  
were selected for comparison because they represent different approaches 
of reducing performance penalties for LSM-tree. 

{\bf WiscKey} was developed for SSD. It employs a value log to store each actual KV pair.
In the memtable and SST files, WiscKey keeps a KV pair's offset in the value log.
As a result, like \DeLSM, WiscKey does not move actual KV pairs. 
However, WiscKey has to manually index KV pairs in the value log. Worse, WiscKey must 
manage the log space by itself. Given a workload with many deletions, its
garbage collection significantly impairs performance. \DeLSM, on the other hand,
delegates such space management to the file system.

{\bf LDS} took an opposite direction to \DeLSM by bypassing file system and
mapping LSM-tree's data directly to a block storage device. Although LDS avoids 
file-level operations and attempts to preserve sequential writes for disk drives,
its compactions still move actual KV pairs across storage space.

{\bf PebblesDB} was inspired by the skip-list data structure. It places `guards' at each level
to partition the key range of the level into disjoint segments.
Such guards help PebblesDB avoid {\em rewriting} the same KV pairs at a level, 
but it still writes them once into the level during a compaction.

{\bf BoLT} is also a contrast to \DeLSM as BoLT creates a single compaction file
for each compaction in order to reduce the calls of {\tt sync}s. 
BoLT uses logical SST indexes
to track KV pairs in the compaction file. 
In a compaction, BoLT physically moves KV pairs between levels and generates
logical indexes though.

\begin{figure*}[t]
\centering
\begin{subfigure}[t]{0.5\columnwidth}
\centering
\includegraphics[width=\columnwidth]{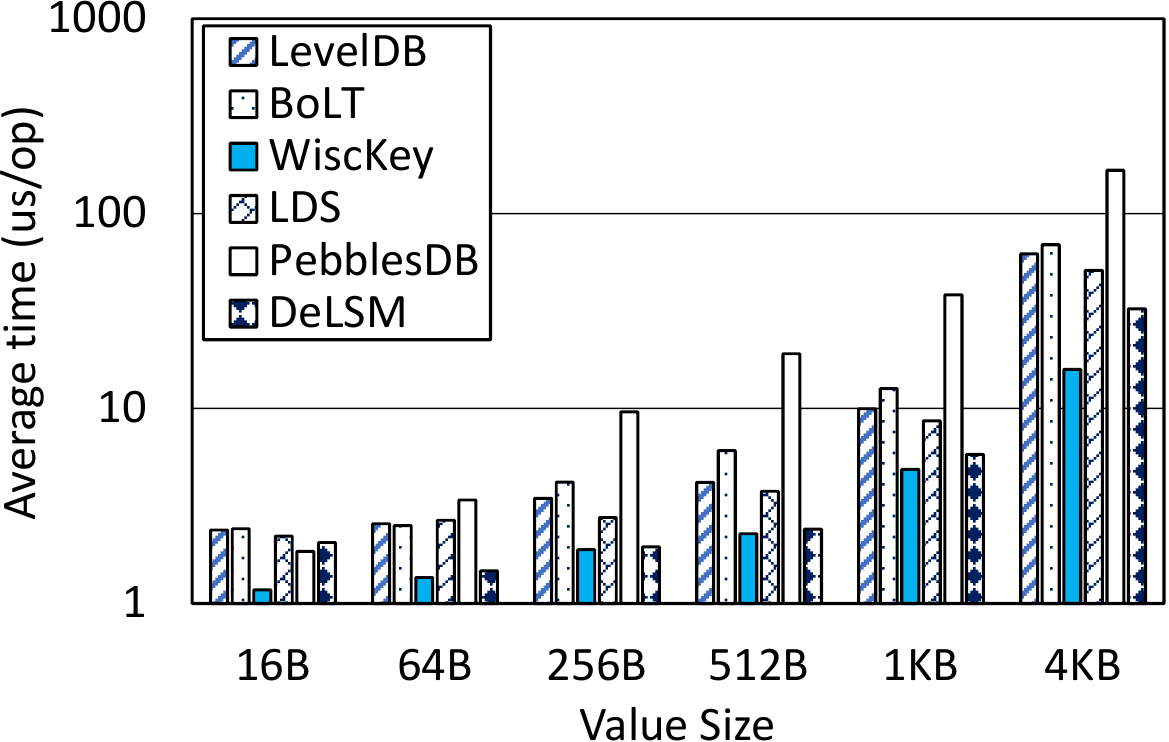}
\caption{\em FillSeq (Sequential Writes)}\label{fig:fillseq}
\end{subfigure}
\hfill
\begin{subfigure}[t]{0.5\columnwidth}
\centering
\includegraphics[width=\columnwidth]{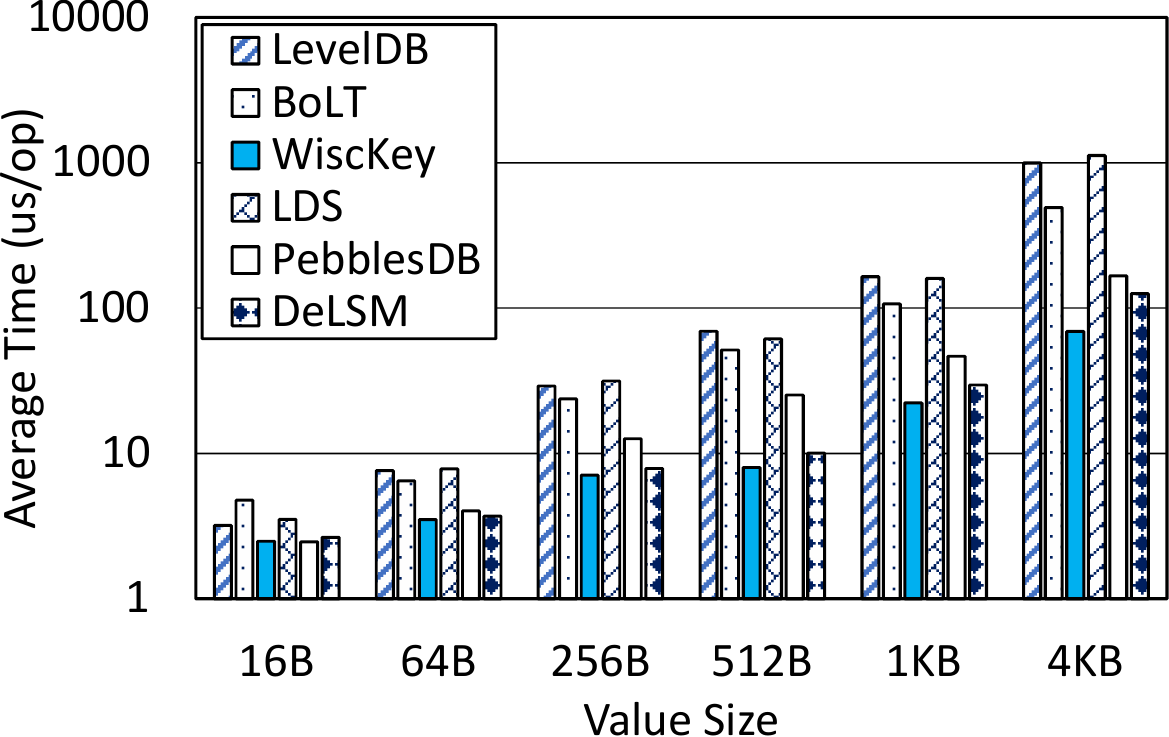}
\caption{\em FillRandom (Random Writes)}\label{fig:fillrandom}
\end{subfigure}
\hfill
\begin{subfigure}[t]{0.5\columnwidth}
	\centering
	\includegraphics[width=\columnwidth]{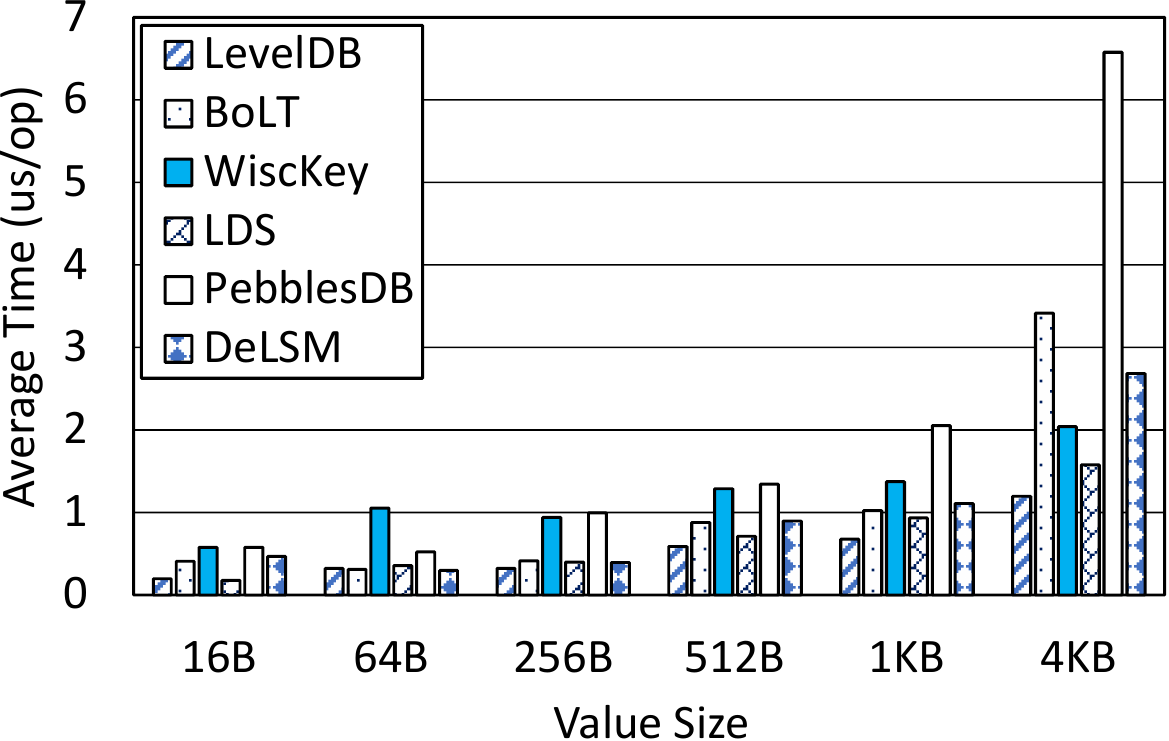}
	\caption{\em ReadSeq (Sequential Reads)}\label{fig:readseq}
\end{subfigure}
\hfill
\begin{subfigure}[t]{0.5\columnwidth}
	\centering
	\includegraphics[width=\columnwidth]{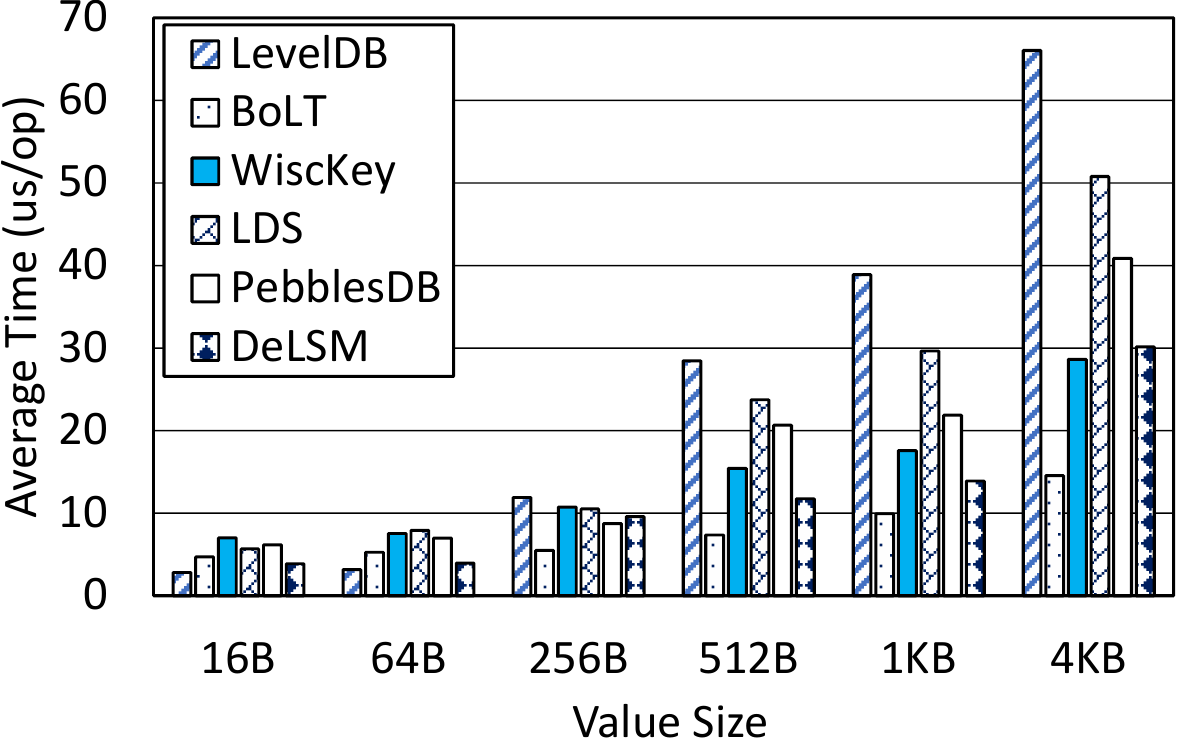}
	\caption{\em ReadRandom (Random Reads)}\label{fig:readrandom}
\end{subfigure}
\caption{A Comparison between Six LSM-tree variants with db\_bench}\label{fig:dbbench}
\end{figure*}

\begin{figure*}[t]
	\centering
	\begin{subfigure}[t]{\columnwidth}
		\centering
		\includegraphics[width=\columnwidth]{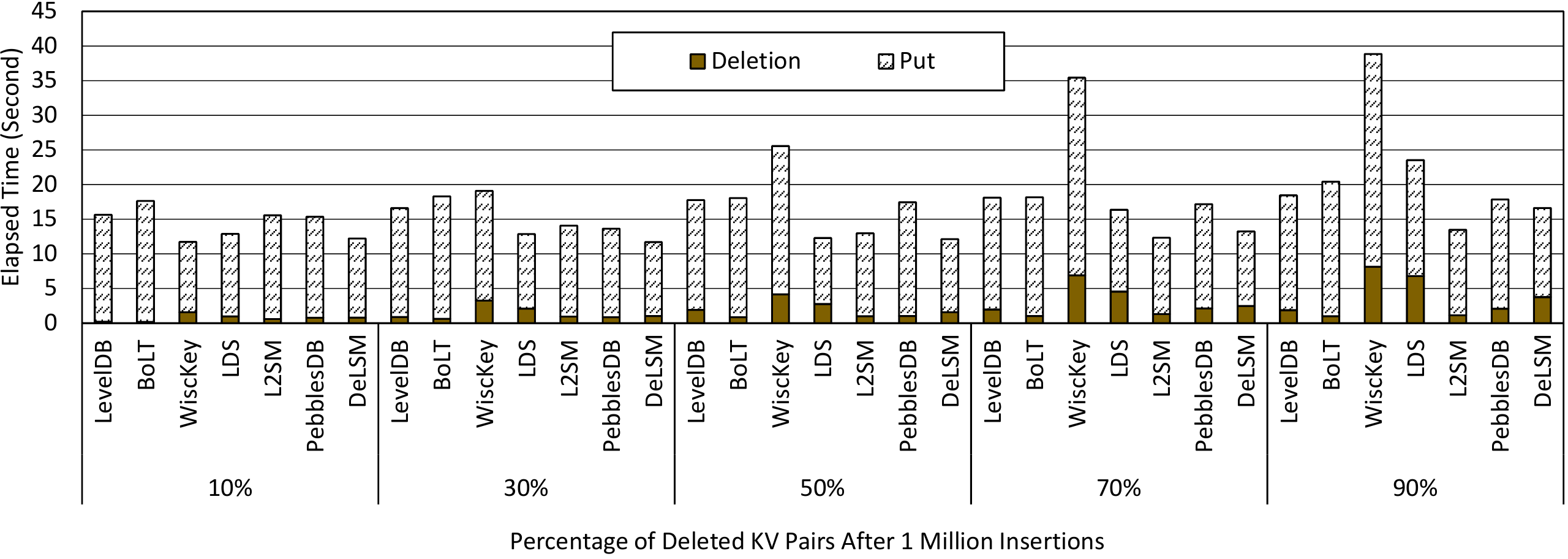}
		\caption{\em Deletion Time and the 2nd Put Time of Write-Delete-Write}\label{fig:Hybrid-write}
	\end{subfigure}
\hfil
	\begin{subfigure}[t]{\columnwidth}
		\centering
		\includegraphics[width=\columnwidth]{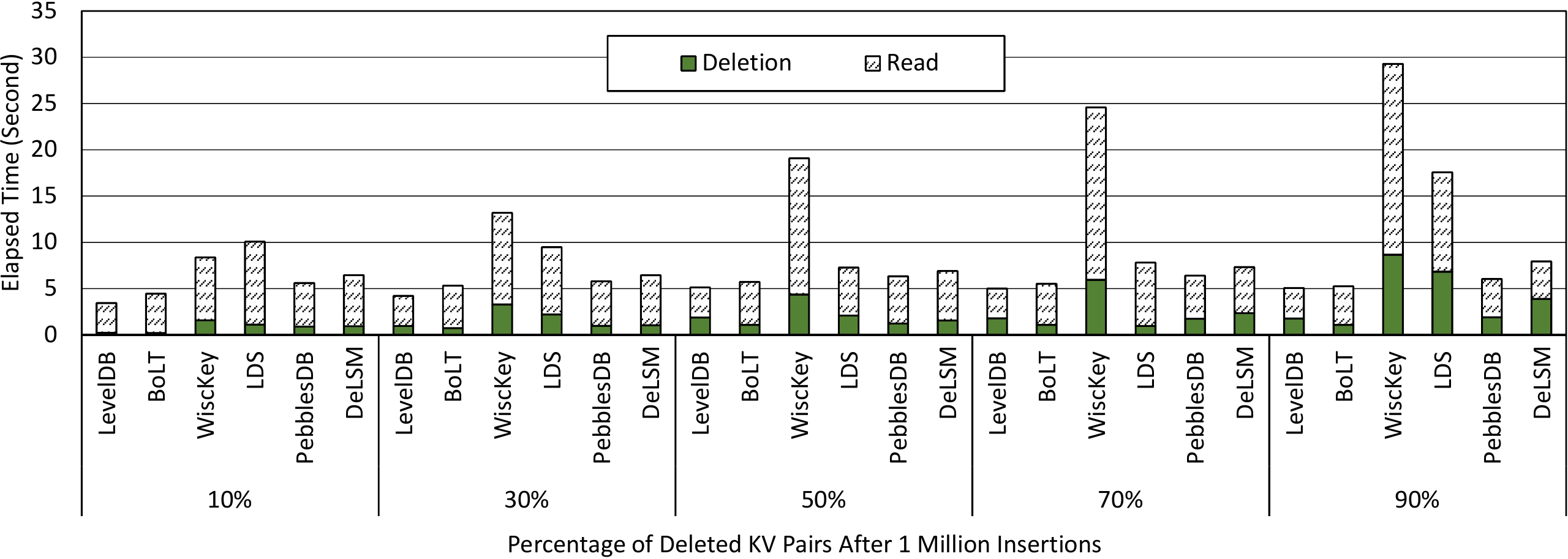}
		\caption{\em Deletion Time and Read Time of Write-Delete-Read}\label{fig:Hybrid-read}
	\end{subfigure}
	\caption{A Comparison between Six LSM-tree variants with Write-Delete-Write and Write-Delete-Read Workloads}\label{fig:hybrid}
\end{figure*}

\subsection{Microbenchmarks}\label{sec:microbenchmarks}

\subsubsection{Sequential and Random Writes without Deletions}\label{sec:normal-write}

We used the default db\_bench as the microbenchmark 
with 16B keys and a variety of value sizes, i.e., 16B, 64B, 256B, 512B, 1KB, and 4KB.
We chose these value sizes because studies had revealed that small values  
are prevalent and even dominant in today's typical real-world workloads~\cite{LSM:LSM-trie:ATC-2015,LSM:Facebook:FAST-2020}.
We collected the average execution time (microsecond/operation) by 
running fillseq (sequential writes), fillrandom (random writes), readseq (sequential reads), and 
readrandom (random reads) of db\_bench
with 10 million KV pairs.
Fillseq (resp. readseq) sequentially puts (resp. searches) KV pairs in ascending order of keys
while fillrandom (resp. readrandom) puts (resp. searches) KV pairs in a random order.

\autoref{fig:fillseq} and~\autoref{fig:fillrandom}
capture the results of write performances of six LSM-tree variants, 
from which we can obtain several observations.
Firstly, we have built \DeLSM on top of LevelDB, and \DeLSM significantly 
boosts the write performance of LevelDB, by up to 18.9$\times$ with fillrandom 
and 4KB values. \DeLSM differs from LevelDB mainly in 1) the reorganization of KV pairs
in the fashion of directory-files and 2)  
the retention of multiple immutable memtable for minor compactions.
The reorganization with SST directories and KV files 
enables \DeLSM to avoid moving KV pairs in the background for major compactions,
thereby incurring much fewer disk I/Os compared to LevelDB.  
Maintaining multiple memtables yet
helps to avoid stalling in the foreground like LevelDB to wait for an ongoing minor compaction.
Consequently, \DeLSM manages to yield much higher performance. 

Secondly, \DeLSM also dramatically outperforms  
other LSM-tree variants except WiscKey with much less execution time in writing
KV pairs, especially with fillrandom (cf.~\autoref{fig:fillrandom}).
Take fillrandom and 4KB value size for illustration again. The average write
time of BoLT, LDS, and PebblesDB is 14.5$\times$, 13.0$\times$
and 6.3$\times$ that of \DeLSM, respectively. In spite of reducing I/O amplifications
in different approaches, these three LevelDB-like LSM-tree variants still have
to physically move KV pairs between levels by their very designs.
Therefore, their write performances are lower than that of \DeLSM which 
only moves {\tt dentry}s in major compactions.

\begin{table*}[ht]
	\centering
	\footnotesize 
	\begin{tabular}{lrrrrrrrrr}		
		\cline{1-9}
		\multicolumn{1}{|c|}{\multirow{2}{*}{LSM-trees}} & \multicolumn{4}{c|}{16B Key, 512B Value}  & \multicolumn{4}{c|}{16B Key, 4KB Value} &  \\ \cline{2-9}
		\multicolumn{1}{|c|}{}                  & \multicolumn{1}{c|}{1st Write} & \multicolumn{1}{c|}{Delete} & \multicolumn{1}{c|}{2nd Write} & \multicolumn{1}{c|}{Overall} & \multicolumn{1}{c|}{1st Write} & \multicolumn{1}{c|}{Delete} & \multicolumn{1}{c|}{2nd Write} & \multicolumn{1}{c|}{Overall} &  \\ \cline{1-9}
		\multicolumn{1}{|c|}{LevelDB}    & \multicolumn{1}{r|}{11.8$\times$}      & \multicolumn{1}{r|}{8.1$\times$}    & \multicolumn{1}{r|}{12.6$\times$}      & \multicolumn{1}{r|}{12.1$\times$}    & \multicolumn{1}{r|}{15.8$\times$}      & \multicolumn{1}{r|}{20.1$\times$}   & \multicolumn{1}{r|}{17.0$\times$}      & \multicolumn{1}{r|}{16.5$\times$}    &  \\ \cline{1-9}
		\multicolumn{1}{|c|}{BoLT}                  & \multicolumn{1}{r|}{12.6$\times$}      & \multicolumn{1}{r|}{10.6$\times$}   & \multicolumn{1}{r|}{13.7$\times$}      & \multicolumn{1}{r|}{13.1$\times$}    & \multicolumn{1}{r|}{16.3$\times$}      & \multicolumn{1}{r|}{50.6$\times$}   & \multicolumn{1}{r|}{17.6$\times$}      & \multicolumn{1}{r|}{17.0$\times$}    &  \\ \cline{1-9}
		\multicolumn{1}{|c|}{WiscKey}                  & \multicolumn{1}{r|}{1.2$\times$}       & \multicolumn{1}{r|}{185.6$\times$}  & \multicolumn{1}{r|}{35.9$\times$}      & \multicolumn{1}{r|}{21.0$\times$}    & \multicolumn{1}{r|}{1.5$\times$}       & \multicolumn{1}{r|}{1471.6$\times$} & \multicolumn{1}{r|}{42.3$\times$}      & \multicolumn{1}{r|}{24.7$\times$}    &  \\ \cline{1-9}
		\multicolumn{1}{|c|}{LDS}    & \multicolumn{1}{r|}{12.7$\times$}      & \multicolumn{1}{r|}{11.5$\times$}   & \multicolumn{1}{r|}{13.0$\times$}      & \multicolumn{1}{r|}{12.8$\times$}    & \multicolumn{1}{r|}{18.7$\times$}      & \multicolumn{1}{r|}{68.1$\times$}   & \multicolumn{1}{r|}{20.0$\times$}      & \multicolumn{1}{r|}{19.5$\times$}    &  \\ \cline{1-9}
		\multicolumn{1}{|c|}{PebblesDB}       & \multicolumn{1}{r|}{8.0$\times$}       & \multicolumn{1}{r|}{8.8$\times$}    & \multicolumn{1}{r|}{7.9$\times$}       & \multicolumn{1}{r|}{8.0$\times$}     & \multicolumn{1}{r|}{12.3$\times$}      & \multicolumn{1}{r|}{52.5$\times$}   & \multicolumn{1}{r|}{12.0$\times$}      & \multicolumn{1}{r|}{12.3$\times$}    &  \\ \cline{1-9}
		\multicolumn{1}{|c|}{\DeLSM}  & \multicolumn{1}{r|}{2.5$\times$}       & \multicolumn{1}{r|}{18.5$\times$}   & \multicolumn{1}{r|}{2.9$\times$}       & \multicolumn{1}{r|}{3.0$\times$}     & \multicolumn{1}{r|}{2.4$\times$}       & \multicolumn{1}{r|}{20.0$\times$}   & \multicolumn{1}{r|}{2.6$\times$}       & \multicolumn{1}{r|}{2.5$\times$}     &  \\ \cline{1-9}
	\end{tabular}
	\caption{The Write Amplifications of LSM-tree Variants with Write (10 million)--Delete (50\%)--Write (10 million) Workloads}\label{tab:ampl}
	\normalsize
\end{table*}

The third observation is on the comparison between \DeLSM and WiscKey, which indicates
that \DeLSM is evidently inferior to WiscKey, particularly with larger values.
For example, with fillrandom 
and 4KB values, \DeLSM is 52.0\% slower than
WiscKey. WiscKey appends all arriving KV pairs
to a value log with disk-favored sequential writes and like \DeLSM,
WiscKey only redeploys the key and offset to the KV in the value log, rather than actual KV pairs in a major compaction. WiscKey also does not
 dump KV pairs through minor compactions. Although \DeLSM uses a log
 to efficiently absorb arriving KV pairs like WiscKey,
it generates individual KV files in minor compactions over time, 
which is likely to
entail random writes as well as {\tt inode} and {\tt dentry} operations~\cite{LSM:LDS:TPDS-2018}. 
Worse, \DeLSM needs more time
to make a KV with larger value into a KV file. 
As a result, \DeLSM cannot compete against WiscKey on workloads that 
are fully composed of write requests.

\subsubsection{Random Writes with Deletions}\label{sec:delete-write}
 
In contrast to \DeLSM that relies on file system for
space management, WiscKey contains a module of garbage collection to
reclaim the space of value log occupied by obsolete data.
In the process of garbage collection, WiscKey scans the value log and 
separates valid KV pairs from deleted ones by moving the former to the tail of value log.
However, db\_bench's fillseq and fillrandom do not issue any deletion request.
Therefore, the garbage collection of WiscKey was never triggered
in the standard tests with db\_bench.
For a fair comparison, we enhanced db\_bench with two more workloads that are both with three
stages. One is write-delete-write that first randomly writes 1 million KV pairs, each of which is with 16B key and 512B values, then deletes a part of them,
and puts 1 million new KV pairs at last. The other one is write-delete-read with the 
 third stage changed to read out all 1 million KV pairs put at the first stage after partial deletions.
 We varied the percentage of KV pairs deleted at the second stage to be 10\%, 30\%, 50\%,
 70\%, and 90\%.
As~\autoref{fig:fillrandom} has presented results with random writes,
~\autoref{fig:Hybrid-write} shows the average execution time
for deletions and the second batch of random writes.
A deletion is lightweight compared to writing a KV pair with 512B value.
However, WiscKey has demanded much longer time than others. 
For example, with 30\% and 70\% deletions, the average deletion time of 
WiscKey is 3.2$\times$ and 2.9$\times$ that of \DeLSM, because WiscKey frequently
executed garbage collections to move valid KV pairs and reclaim space.
In the subsequent writing of new KV pairs after 30\% and 70\% deletions,
the average write time of WiscKey is 1.5$\times$ and 2.7$\times$ that of \DeLSM, respectively.
During the third stage,  
WiscKey was still doing garbage collections. With more KV pairs deleted
in the second stage (30\% to 70\%), it should encounter more obsolete 
data and perform more extra I/Os for garbage collection. This explains why
 WiscKey is interestingly slower than \DeLSM in the latter two stages of deleting and
 writing KV pairs
and the performance gap between them becomes wider with more KV pairs deleted.

In addition, for \DeLSM, deletions need file system to do with 
{\tt dentry}s and {\tt inode}s.
Such cost, albeit being little, impairs performance, especially considering
LevelDB-like LSM-tree variants
only put a key with a deletion tag.
Such is the reason why the average deletion time of \DeLSM is longer
than some LSM-tree variants.

\subsubsection{Write Amplifications}

The primary target of \DeLSM is to reduce performance penalties caused by write amplifications.
To make a quantitative analysis of the aforementioned write performances,
we have collected the actual write I/Os through {\tt iotop}\footnote{\url{https://www.man7.org/linux/man-pages/man8/iotop.8.html}}
when running write-delete-write workloads. 
\autoref{tab:ampl} captures the amplification factors of actual data size committed to disk 
against data size that users issued with six LSM-tree variants. 
Firstly, the results of writing 10 million KV pairs, deleting half of them, and then
writing new 10 million KV pairs align with the observations of write performances we have discussed in 
Section~\ref{sec:normal-write} and Section~\ref{sec:delete-write}.
For example, with 4KB values written in the first stage, \DeLSM writes 2.4$\times$ data
to disk, including log entries, data and metadata ({\tt inode}s and {\tt dentry}s) for KV files, and manifest records. 
However, compared to other LSM-tree variants except WiscKey, such an amplification factor is much lower
as they caused 12.3$\times$ to 18.7$\times$ write I/Os, particularly due to redeploying KV pairs in compactions. 
These numbers endorse why \DeLSM is superior in write performances and in turn
justify the soundness of \DeLSM that leverages file system's directory management
to avoid moves of KV pairs.
In the meantime, 
WiscKey incurred 1.5$\times$ write I/Os in putting down 10 million KV pairs with 4KB values,
37.5\% fewer than that of \DeLSM. 
This margin and the speed mismatch between WiscKey's sequential writes and \DeLSM's random 
writes justify their performance gap (52.0\%) reported in Section~\ref{sec:normal-write}.

When handling deletions and subsequent write requests, 
however, the module of garbage collection would badly
move valid KV pairs for WiscKey, which amplified the write I/Os by as much
as 1,471.6$\times$ and 42.3$\times$, respectively, with 4KB values.
These two are much higher than those factors of other LSM-tree variants.
Nonetheless, because WiscKey manages the space of value log by itself,
the functionality of space recycling and defragmentation is essential but costly. 
Comparatively, \DeLSM embraces higher flexibility and efficacy as it
 delegates the underlying file system to manage the space allocation and deallocation
for SST directories and KV files,

\vspace{2ex}
To summarize the tests for write performance, 
like all scientific designs, \DeLSM is greatly effectual for some workloads
but not very well for some others.
It drastically
outperforms LevelDB-like LSM-tree variants
that move actual KV pairs for compactions with much fewer disk I/Os.
As to \DeLSM and WiscKey that also does not move actual KV pairs, 
they both exhibit respective strengths and weaknesses
in serving different workloads 
while \DeLSM is more flexible without application-level space management.
 
\subsubsection{Read Performance}

\autoref{fig:readseq} and \autoref{fig:readrandom} show the read performances
of six LSM-tree variants with readseq (sequential reads) and readrandom (random reads), respectively. 
Note that the Y axes of these three 
diagrams are not in logarithmic scale.
From them we can observe that 1) for sequential reads that resemble range queries,
\DeLSM achieves comparable or even faster performance compared to other LSM-tree variants,
especially with smaller values, and 2) for random reads that emulate point searches,
\DeLSM is generally faster than the most of other LSM-tree variants. 
With sequential reads, \DeLSM is inferior as its values are stored in separate 
files and cannot efficiently generates sequential accesses onto disk drive.
As to random reads, however, the meta file and the {\tt dentry} management of file system
enable \DeLSM to quickly locate a KV file. For instance,
with readrandom and 1KB values, the performance of \DeLSM is 2.8$\times$
that of LevelDB.

\section{Related Works}\label{sec:related}

Researchers have considered how to  
reduce I/O amplifications caused by compactions within LSM-trees~\cite{LSM:bLSM:SIGMOD-2012,LSM:LSM-trie:ATC-2015,LSM:WiscKey:FAST-2016,LSM:TRIAD:ATC-2017,LSM:LSbM-tree:ICDCS-2017,LSM:LWC-tree:TOS-2017,LSM:PebblesDB:SOSP-2017,LSM:SlimDB:VLDB-2017,LSM:FlashKV-OpenSSD:TECS-2017,LSM:LDC-SSD:ICDE-2019,LSM:FPGA-accelerated-compaction:FAST-2020,LSM:FPGA-based-compaction:ICDE-2020,LSM:GearDB:ATC-2020,LSM:L2SM:ICDE-2021,LSM:Nova-LSM:SIGMOD-2021}.
However, their designs mainly followed the conventional way of grouping multiple 
KV pairs in one integral SST file.
Shetty~et al.~\cite{LSM:KVFS:FAST-2013} proposed VT-Tree that attempts to reduce disk reads and writes by merging sorted segments of non-overlapping levels in the tree. 
Similarly, Wu et al.~\cite{LSM:LSM-trie:ATC-2015} proposed LSM-trie that uses a trie (prefix tree) to organize KV pairs in an LSM-tree and compact SST files without
overlapping keys.
Both of them still move excessive KV pairs in a compaction.
Balmau et al.~\cite{LSM:TRIAD:ATC-2017} and Huang et al.~\cite{LSM:L2SM:ICDE-2021}
considered how to separate hot and cold data so as to reduce frequently
moving KV pairs caused by hot ones in a compaction. 
The aforementioned PebblesDB~\cite{LSM:PebblesDB:SOSP-2017} 
fragments an SST file into fine-grained pieces and appends such pieces 
to the next level during a compaction. In spite of writing KV pairs to a level only once, the move of KV pairs is inevitable. 
\DeLSM, nonetheless, redeploys {\tt dentry}s pointing to KV pairs (files) in compactions,
thereby gaining high efficiency. Also, 
\DeLSM does not employ complex structures to index or separate data,
but relies on the underlying file system.

\DeLSM can also be viewed as an approach that leverages file system to accelerate LSM-trees, which
is a stark contrast to research works that use databases, especially LSM-trees,  
to augment and accelerate local or distributed file systems~~\cite{LSM:KVFS:FAST-2013,LSM:TableFS:ATC-2013,FS:BetrFS-TokuDB:FAST-2015,LSM:IndexFS:SC-2014,LSM:LocoFS:SC-2017,LSM:SlimDB:VLDB-2017,LSM:DeltaFS:SC-2018,LSM:KVSSD:DATE-2018,LSM:FS-SSD:OSDI-2021}.
For example, Shetty~et al.~\cite{LSM:KVFS:FAST-2013} utilized the aforesaid VT-Tree to 
build KVFS that translates an access request to one or more KV operations
for VT-Tree to process. Ren, Gibson, and their collaborators successively developed TableFS, IndexFS, and SlimDB that make use of LSM-tree to manage
file system metadata and small files, respectively, in local and cluster file systems~\cite{LSM:TableFS:ATC-2013,LSM:IndexFS:SC-2014,LSM:SlimDB:VLDB-2017}.
Li et al.~\cite{LSM:LocoFS:SC-2017} reshaped the 
organization and management of directory and file metadata with LSM-tree
for their LocoFS.
Zheng et al.~\cite{LSM:DeltaFS:SC-2018} proposed to leverage a customized LSM-tree
in their DeltaFS to index and manage data 
for fast queries.
 
\DeLSM is a generic, software-only solution while
there are LSM-tree variants that have either taken into account the characteristics of storage devices or made use of specific hardware supports~\cite{LSM:WiscKey:FAST-2016,LSM:Tucana:CPU-overhead:ATC-2016,LSM:FlashKV-OpenSSD:TECS-2017,LSM:NoveLSM:ATC-2018,LSM:LDS:TPDS-2018,LSM:LDC-SSD:ICDE-2019,LSM:FPGA-accelerated-compaction:FAST-2020,LSM:FPGA-based-compaction:ICDE-2020,LSM:MatrixKV:ATC-2020}. 
The emergence of SSDs motivated the designs of SSD-conscious LSM-trees, such as
WiscKey~\cite{LSM:WiscKey:FAST-2016}, FlashKV~\cite{LSM:FlashKV-OpenSSD:TECS-2017}, and LDC~\cite{LSM:LDC-SSD:ICDE-2019}. 
Using KV-store in turn to redesign SSDs was also exploited~\cite{LSM:KV-SSD:HPCA-2017,LSM:KV-SSD:Systor-2019,LSM:in-storage-KV-PinK:ATC-2020,LSM:in-SSD-index:OSDI-2021}. Meanwhile, 
Yao et al.~\cite{LSM:LWC-tree:TOS-2017, LSM:GearDB:ATC-2020} developed LSM-trees for the Shingled Magnetic Recording (SMR) disk drives while Kannan et al.~\cite{LSM:NoveLSM:ATC-2018}, Kaiyrakhmet et al.~\cite{LSM:SLM-DB:FAST-2019}, and Yao et al.~\cite{LSM:MatrixKV:ATC-2020} devised LSM-tree variants with byte-addressable non-volatile memories (NVM). 
Reducing the CPU overheads for LSM-tree~\cite{LSM:Tucana:CPU-overhead:ATC-2016} and transfiguring LSM-tree into in-memory database~\cite{LSM:Accordion:VLDB-2018} were considered as well.
Recent studies~\cite{LSM:FPGA-accelerated-compaction:FAST-2020,LSM:FPGA-based-compaction:ICDE-2020}
 sought aid from FPGA for hardware supports to accelerate compactions. 
Comparatively, the idea of \DeLSM is more portable and applicable.

\section{Conclusion}\label{sec:conclusion}

We revisit LSM-tree and attempt to reduce performance penalties
caused by major compactions through making a KV pair into
an individual KV file, no longer a component of one large SST file.
As a result, we transform a compaction of physically moving actual KV pairs
into a series of lightweight {\tt dentry} operations that link and unlink 
KV files. This is the essence of \DeLSM proposed in this paper.
Preliminary experiments confirm that a prototype of \DeLSM 
significantly outperforms LevelDB, one that the prototype has been built on,
by up to 18.9$\times$ and 2.8$\times$ on write and read performances,
respectively. \DeLSM also shows superior performance compared against state-of-the-art LSM-tree
variants in different dimensions. Currently
we are working on the optimization of \DeLSM and 
conducting furthermore tests.


\bibliographystyle{unsrt}
\bibliography{tree}

\end{document}